\documentclass[aps,showpacs,twocolumn]{revtex4}
\usepackage{amsmath, amsthm,times}
\usepackage{amssymb}
\usepackage{graphicx}

\begin{document}

\sloppy

\title{Verifying multi-partite mode entanglement of W states}

\author{Pavel Lougovski$^{1}$\footnote{present address: Department of Radiation Oncology, \\
Stanford University School of Medicine, \\
Stanford, CA 94305-5847}, S.J. van Enk$^{1,2}$,
Kyung Soo Choi$^3$, Scott B. Papp$^3$, Hui Deng$^3$, H.J. Kimble$^3$}

\address{
$^{1}$Department of Physics and Oregon Center for Optics, 1274 University of Oregon, Eugene, OR 97403\\
$^{2}$Institute for Quantum Information, California Institute of Technology, Pasadena, CA 91125\\
$^3$ Norman Bridge Laboratory of Physics, California Institute of Technology, Pasadena, CA 91125}

\begin{abstract}
We construct a method for verifying mode entanglement of $N$-mode W states. The ideal W state contains exactly one excitation symmetrically shared between $N$ modes, but our method takes the existence of higher numbers of excitations into account, as well as the vacuum state and other deviations from the ideal state. Moreover, our method distinguishes between full $N$-party entanglement and 
states with $M$-party entanglement with $M<N$, including mixtures of the latter. We specialize to the case $N=4$ for illustrative purposes. In the optical case, where excitations are photons, our method can be implemented using linear optics.
\end{abstract}

\pacs{03.65.Mn, 42.50.Ex}

\maketitle

\section{Introduction}

Detecting and classifying entanglement is an important challenge in the field of quantum information science. One problem is of a theoretical nature, to decide whether a given density matrix $\rho$ of multiple quantum systems is entangled or separable. Even for bipartite systems, this is a hard problem for which no efficient general solution is known for higher-dimensional Hilbert spaces, although a simple test based on the negativity of the partial transpose of the density matrix leads to a sufficient criterion for entanglement \cite{Witness}. If $\rho$ is entangled, the next issue is how to classify the type of entanglement.  For more than two subsystems, the full classification of all entanglement  classes is as yet an unsolved problem (see, e.g., Refs.~\cite{W,class}).

In an experiment the practical task of detecting entanglement is even harder. If one would perform a full tomographic measurement, then in the limit of infinitely many data one would end up with an arbitrarily accurate estimate of a density matrix $\rho$, and thus reduce the experimental problem to the theoretical problem mentioned above. In all other cases, 
one needs different tests that make use of less than full knowledge of the density matrix. 
The main practical disadvantage of full tomography is the rapidly growing number (with the number of quantum systems and with the dimension of the Hilbert spaces involved) of measurements needed to find all elements of a density matrix. The other challenge is obtaining a physical density matrix from a {\em finite} set of measured data~\cite{tomo,Blume-Kohout}.    

Thus there is an evergrowing demand for simpler experimental tests revealing entanglement. Fortunately, for bipartite systems there exists a handful of different experimental techniques for entanglement detection~\cite{vanEnkKimbleLutkenhaus,review}.  Here we will focus on a particular type of multi-partite entangled states (namely, W states \cite{W}) that can be produced in systems with variable numbers of excitations. We think in particular of experiments with atomic ensembles (see, e.g. \cite{ensembles}) based on the DLCZ protocol \cite{dlcz} in which information can be stored in the number of atomic excitations of each ensemble, as well as of experiments on photonic systems (see, e.g. \cite{photon}), where the number of photons in a given mode can be used as a quantum variable. In the following we will use the words ``excitation'' and ``photon'' interchangeably. 

We define the state $|W\rangle$ as a mode-entangled analogue of standard $N$-partite W states of qubits. It is a pure state where a single excitation is shared symmetrically between $N$ modes \footnote{Following convention, we set all phase factors equal to unity; our entanglement detection method, however, will not make any assumption about the phase factors of the state actually generated in one's experiment, see Eq.~(\ref{Wbasis})}
\begin{equation}
\label{Nmode}
|W\rangle = \frac{1}{\sqrt{N}}\sum\limits^{N}_{i=1} |0,\cdots,0_{i-1},1_{i},0_{i+1},\cdots,0 \rangle, 
\end{equation}
where $|0\rangle$ denotes a state of a mode with no excitations and $|1\rangle$ is a state with a single excitation. The subscripts $i=1\ldots N$ refer to  modes that are in {\em  spatially distinct locations}, so that the concepts of ``local operations'' and hence entanglement are unambiguously defined \cite{enk}. 

We solve the problem of detecting the entanglement of a W state (and it noisy cousins) in two steps. In Section II we will focus on detecting and classifying entanglement within the subspace of a fixed {\em total} number of excitations (in all modes together), namely one. In Section III we complete the analysis by including the remaining parts of Hilbert space, the subspace with no excitations and the subspace with more than a single excitation in total. Including both subspaces is crucial in the analysis. Earlier detection schemes \cite{Nha} for W states in the context of photons were incomplete due to the neglect of states with multiple photons. Moreover, we will discuss how to include imperfections such as losses, most releveant for the actual implementation of our method \cite{exp}.

\section{Genuine $N$-mode one-photon entanglement}
$N$ parties can be entangled in many different ways.
In some papers ``genuine'' $N$-party entangled states include states that are mixtures of $M$-party entangled states with $M<N$, as long as such mixtures are not biseparable along any particular splitting of the $N$ parties into two groups (for instance, \cite{loock}). Here, however, we will classify such mixtures as $M$-party entangled states, and the name ``genuine $N$-party entanglement'' in our case is reserved for states that can only be written  as a mixture of pure states that all possess $N$-party entanglement. Thus our criterion for genuine $N$-party entanglement is  more severe.

Recently it has been suggested that uncertainty relations can be used as an entanglement criterion for finite-dimensional systems (~\cite{HofmannTakeuchi}; see also \cite{Wang} for an experimental implementation using local observables on two qubits). The uncertainty principle sets up a fundamental limit on how accurately observables of a quantum system can be simultaneously determined.  For instance, if $\{M_{i}\}, i=1\ldots K$ is some set of observables, then the measurement uncertainty in a given state $\rho$ is given by the sum of variances of all observables $M_{i}$, i.e., $\sum_i \delta M_i(\rho)^2$. This sum equals zero if and only if the state for which measurements of all  $M_{i}$ are performed is a simultaneous eigenstate of all $M_{i}$. If there is no such state (when the observables are not all mutually commuting) then there exists a positive number $C$ such that
\begin{equation}\label{variance}
\sum\limits^{K}_{i=1} \delta M_i(\rho)^{2} \geq C.
\end{equation}
Hofmann and Takeuchi pointed out that the existence of the lower uncertainty bound $C$ can be employed as a separability criterion \cite{HofmannTakeuchi}. Indeed, if for some fixed set of observables an inequality of the form (\ref{variance}) holds for{ \it all} separable states then its violation is a signature of entanglement.  

The uncertainty bound has another obvious but important property. Namely, one can never decrease the average uncertainty by mixing different states. In other words, for any state $\rho=\sum_{m}p_{m}\rho_{m}$ and any observable $A$ the following inequality holds
\begin{equation}\label{bound}
\delta A(\rho)^{2}\geq \sum_{m}p_{m}\delta A(\rho_m)^{2}.
\end{equation}
The proof is rather straightforward and can be found in \cite{HofmannTakeuchi}. 

With the uncertainty criterion at hand we still have some flexibility over the type of observables to choose. In principle, all observables can be divided into two groups -- local and nonlocal. Whereas local observables can be measured separately for each and every party and therefore tend to be easier to access in an experiment, they often cannot reliably detect genuine multipartite entanglement. Nonlocal observables, on the other hand, require a simultaneous nonlocal measurement of several parties at a time, which often is experimentally challenging. Here we show how experimentally accessible nonlocal observables can be constructed to unambiguously detect genuine multipartite entanglement of the W-type.

The basic idea behind the construction of nonlocal observables is to choose them as projectors onto a basis of $N$-partite entangled states. Simultaneous eigenstates of these projectors are necessarily entangled states, and the variance in the projectors is minimized for  $N$-party entangled states. A sufficiently small variance is then a sufficient criterion for genuine $N$-party entanglement.  In order to illustrate this idea we will consider a system of four modes sharing a single photon. The problem at hand is then to find a set of nonlocal observables which allows to separate all four-mode separable and biseparable states from the genuinely four-mode entangled states such as the W state of Eq.~(\ref{Nmode}). We note that the general construction for an arbitrary $N$  can be done in a similar fashion. Moreover, the nonlocal observables for single photons we use can be measured just using linear optics (beam splitters) and non-number resolving photodetectors.

\begin{figure}
  \includegraphics[scale=0.3]{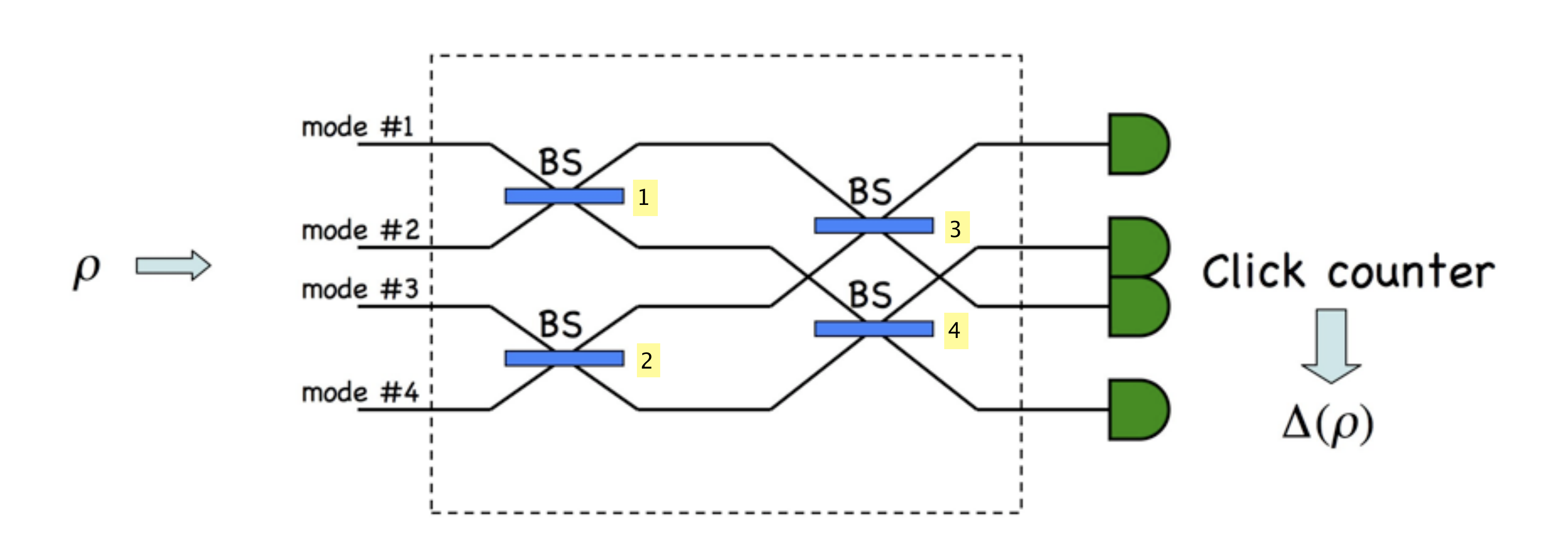}
  \caption{Beam-splitter setup to project onto
  the four W states (\ref{Wbasis1}-\ref{Wbasis4}): 4 input modes are converted into 4 output modes by 4 50/50 lossless beamsplitters, numbered 1--4. From the count statistics of (ideal) detectors placed at the 4 output modes one obtains the quantity $\Delta(\rho)$ defined in Eq. (\ref{measure}).
  Losses, asymmetries in the beamsplitters, and non-ideal detectors are discussed in Section \ref{losses}.}\label{setup}
\end{figure}

The Hilbert space of a system of four modes sharing exactly one photon is spanned by four basis product vectors $\{|1000\rangle,|0100\rangle,|0010\rangle,|0001\rangle\}$. This basis can always be rotated to a basis constituted by four  W-like states,  
\begin{eqnarray}\label{Wbasis1}
|W_{1}\rangle & = & \frac{1}{2}(|1000\rangle + |0100\rangle + |0010\rangle + |0001\rangle), \\
|W_{2}\rangle& = & \frac{1}{2}(|1000\rangle -  |0100\rangle -  |0010\rangle + |0001\rangle), \\
|W_{3}\rangle & = & \frac{1}{2}(|1000\rangle + |0100\rangle - |0010\rangle - |0001\rangle), \\ \label{Wbasis4}
|W_{4}\rangle & = & \frac{1}{2}(|1000\rangle - |0100\rangle + |0010\rangle - |0001\rangle).  
\end{eqnarray}
The mode transformation from the four product states to these four W-like states can be easily decomposed in terms of unitary operations that can be implemented with beamsplitters and phaseshifters (see Figure \ref{setup}).  

The next step is to choose four projectors onto the basis Eqs.~(\ref{Wbasis1}-\ref{Wbasis4}) as nonlocal observables $M_{i}=|W_{i}\rangle \langle W_{i}|, i=1,4$.  
Clearly, the only simultaneous eigenstates of all four operators $M_{i}$ are the four states $|W_{i}\rangle $. The total variance of all $M_{i}$'s vanishes for any one of the states $|W_{i}\rangle $.  In contrast, for product states the total variance is bounded from below, since there exists no simultaneous product eigenstate of all the $M_{i}$. Therefore, we can write down an uncertainty-based entanglement criterion using nonlocal observables for any state $\rho_1$ within the subspace of a single excitation, in terms of the sum of variances of $M_{i}$,
\begin{eqnarray}\label{measure}
\Delta(\rho_1)& =& \sum\limits^{4}_{i=1}{\rm Tr}(\rho_1\left[
|W_{i}\rangle \langle W_{i}|\right]^2)-[{\rm Tr}(\rho_1|W_{i}\rangle \langle W_{i})]^{2}\nonumber\\
&=&
\sum\limits^{4}_{i=1}\left[ \langle W_{i}|\rho_1 |W_{i}\rangle- \langle W_{i}|\rho_1 |W_{i}\rangle^2\right]\nonumber\\
&=&1-\sum_{i=1}^4 \langle W_{i}|\rho_1 |W_{i}\rangle^2, 
 \end{eqnarray}
where the subscript 1 is there to remind us the state contains exactly 1 excitation.

\begin{figure}
  \includegraphics[scale=0.4]{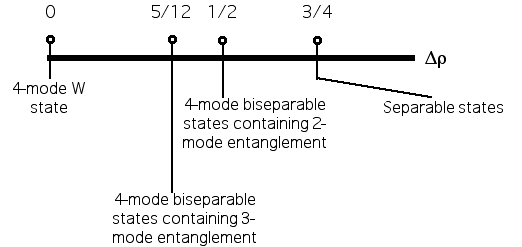}
  \caption{Minimum variances $\Delta$ for various types of four-mode states containing exactly one excitation (photon). }\label{Ent}
\end{figure}

To find the lower bound on $\Delta$ for unentangled states it is sufficient to consider pure states thanks to Eq.~(\ref{bound}). For a pure state $\rho_1=|\alpha\rangle\langle \alpha|$ we have 
\begin{equation}
\Delta(\rho_1) = 1 - \sum\limits^{4}_{i=1} |\langle W_{i}|\alpha\rangle|^{4}.
\end{equation}
The next step is to find the minimum of $\Delta(\rho_1)$ by maximizing $\sum_i |\langle W_{i}|\alpha\rangle|^{4}$ over all separable states $|\alpha\rangle$ containing a single excitation.  There are three types of pure four-mode states that are {\em not} four-mode entangled \footnote{Note that there is only one class of four-mode entangled states with one excitation: states of the W-type $a|0001\rangle+b|0010\rangle+c|0100\rangle+d|1000\rangle$. Our method can be used to detect {\em any} four-mode entangled state within the subspace of a single excitation, by modifying the projectors appropriately.}
\begin{enumerate}
\item The fully separable pure states, which are products of four single-mode states. There are only four such states within the subspace of interest, namely $|1000\rangle$, $|0100\rangle$, $\ldots |0001\rangle$. 
\item Biseparable states with at most two-mode entanglement. Here two modes must be in the vacuum state, and the most general pure state in this class is of the form
$|00\rangle\otimes (a|01\rangle+b|10\rangle)$, or similar states resulting from permuting the different modes.
\item Biseparable states with at most three-mode entanglement. Here at least one mode is in the vacuum state, and the most general pure state, up to permutations of modes, is of the form
$|0\rangle \otimes(a|001\rangle+b|010\rangle +c|100\rangle)$.
\end{enumerate}  
Given the most general pure state within each class it is straightforward to calculate the three corresponding minimum values of $\Delta(\rho_1)$, and the results are depicted in Fig.~\ref{Ent}. For example, for any pure fully separable state $|\alpha\rangle$ the overlap $|\langle W_i|\alpha\rangle|^2=1/4$ for any $i$, and so $\Delta(\rho_1)=3/4$. For general mixtures of fully separable states, this number gives the best possible lower bound on the variance. 
We note that the numerical results from the next Section will confirm the results of Fig.~\ref{Ent}.

As an example, consider the Werner-like mixture of a W state and the maximally mixed state of four modes with a single excitation, $\rho_{mm}=(|0001\rangle\langle 0001|+
|0010\rangle\langle 0010|+|0100\rangle\langle 0100|+|1000\rangle\langle 1000|)/4$,
\begin{equation}\label{wernerlike}
\rho_1(p)=p|W_1\rangle\langle W_1|+(1-p) \rho_{mm}.
\end{equation}
Using the above criterion for $\Delta(\rho_1(p))=3/4-3p^2/4$, we find that for $p>2/3$ we can detect genuine four-mode entanglement, and for $p>\sqrt{3}/3\approx 0.577$ we detect at least three-mode entanglement. Moreover, for any $p<1$, the state $\rho_1(p)$ is entangled, even if just two-mode entangled.

If the number of modes $N$ is arbitrary, then the minimum of $\Delta(\rho_1)$ for biseparable $(N-1)$-mode entangled states can be shown, after some algebra, to be given by $(2N-3)/N(N-1)$. In the limit of large $N$, this bound rapidly approaches zero, hence making it practically impossible to distinguish in this way genuine $N$-partite entanglement from mere $(N-1)$-partite entanglement. 

 Finally, we note similar uncertainty-based entanglement criteria can in principle be applied to all types of $N$-mode states with some fixed total number of excitations. If the total photon number is larger than 1, however, the unitary transformation from product states to a basis consisting of entangled states can in general not be performed with linear optical operations only. Therefore, measurements in such a  basis would no longer be deterministic in that case.
 \section{Detecting W states in an experiment}
Due to experimental imperfections, an actual state, produced in a laboratory, is never a pure state with a fixed number of excitations, such as, say, Eq.~(\ref{Wbasis1}). In experiments with atomic ensembles a state $\rho_W$ is routinely generated whose single-excitation part has a large overlap with a W state, but which contains a significant contribution from the vacuum and from states with more than one excitation. As a conservative estimate, we can ignore coherent superpositions of states with different numbers of excitations [one can get rid of such coherences by local operations, see \cite{kimble}], and hence we can write down $\rho_W$ in a generic form
\begin{equation}\label{density}
\rho_W = p\rho_0 + q\rho_{1} + (1-p-q)\rho_{\geq2},
\end{equation}
where the subscript indicates the number of excitations.
Typically, the magnitude of $1-p-q$ is of the order of 1\% 
or even less. The main source of contamination to the desired single-excitation part is the vacuum. Moreover,  $\rho_1$ is not necessarily a pure state, and is not necessarily a state of $N$ {\em single} modes either. 

Even if the uncertainty measure from the preceding Section would identify the presence of four-mode entanglement in the state $\rho_{1}$, this does not guarantee that $\rho_W$ itself carries any entanglement. The standard counterexample \cite{comment} is a four-mode state of the (unnormalized) form
\begin{equation}
|+\rangle\propto(|0\rangle +\epsilon |1\rangle)^{\otimes 4},
\end{equation}
for which the one-excitation part is genuinely four-party entangled, although the state itself is fully separable.
Therefore, in order to justify the presence of entanglement in an experiment it is not sufficient to measure only the variance $\Delta(\rho_1)$ of the single-excitation part of the density matrix, but it is crucial to additionally measure  the numbers $\{p, q\}$. Once $p,q$ and $\Delta(\rho_1)$ are measured one can check if there exists a completely separable or biseparable state $\rho_{test}$ with the same values of $p,q,\Delta(\rho_1)$. If no such state exists, then one can correctly conclude that $\rho_W$ is entangled. 

More precisely, for fixed values of $p$ and $q$ we wish to find the minimum possible value for the variance $\Delta$, $\Delta_{{\rm min}}$, consistent with the various sorts of biseparable or fully separable states.  In the following we will plot results for the case where $q = 0.1$, which is the relevant case for the experiment \cite{exp}. We will find $\Delta_{{\rm min}}$ in that case as a function of $r:=1-p-q$. 

Before discussing in turn the various classes of separable and biseparable states, we make several remarks:

We note that $\Delta_{{\rm min}}$ within each such class cannot increase with decreasing $q$. The reason is that given any state $\rho$  we can always mix in the vacuum $\rho_0$ without changing the variance $\Delta$, and without increasing the entanglement. But this mixing operation clearly does decrease $q$. Hence $\Delta_{{\rm min}}$ cannot increase with decreasing $q$.

Similarly, we could mix in a fully separable state containing more than a single photon in some given mode, e.g., a tensor product of the vacuum and one mode with 2 or more photons. This again does not affect $\Delta$, and does not increase entanglement, but does decrease $q$. For this reason, in our attempts to find the minimum variance we do not have to consider
states with more than a single photon in any given mode, as those states will have a larger value of $\Delta$ than the minimum possible for given $q$.

Moreover, we could take a state of $N$ single modes and convert it into a state of multiple modes in $N$ locations, by locally applying a random unitary operation. This local operation does not move a state up the entanglement hierarchy and does not affect any of the quantities $\Delta$, $q$ and $r$. Thus, excluding fully separable states and biseparable states of $N$ {\em single} modes is sufficient
for detecting entanglement.

Because $\rho_1$ is subnormalized to $q$, we have,
instead of the inequality (\ref{bound}), the inequality 
\begin{equation}\label{boundq}
q\Delta(\rho_1) \geq \sum\limits^{}_{m=1}p_{m}q_m\Delta(\rho_{m,1}),
\end{equation}
where $q=\sum_m p_m q_m$ and $\rho=\sum_m p_m\rho_m$.

Finally, instead of projecting onto the 4 states (\ref{Wbasis1})--(\ref{Wbasis4}), in an experiment one would really project onto 4 states of the form
\begin{eqnarray}\label{Wbasis}
|W'_{1}\rangle & = & \frac{1}{2}(|1000\rangle +e^{i\phi_1} |0100\rangle + e^{i\phi_2}|0010\rangle +e^{i\phi_3} |0001\rangle), \nonumber\\
|W'_{2}\rangle& = & \frac{1}{2}(|1000\rangle -  e^{i\phi_1}|0100\rangle -  e^{i\phi_2}|0010\rangle +e^{i\phi_3} |0001\rangle), \nonumber\\
|W'_{3}\rangle & = & \frac{1}{2}(|1000\rangle +e^{i\phi_1}|0100\rangle - e^{i\phi_2}|0010\rangle -e^{i\phi_3} |0001\rangle), \nonumber\\ 
|W'_{4}\rangle & = & \frac{1}{2}(|1000\rangle - e^{i\phi_1}|0100\rangle + e^{i\phi_2}|0010\rangle - e^{i\phi_3}|0001\rangle),\nonumber\\
\end{eqnarray}
and vary over all three phases $\phi_k, k=1,2,3$ (which one accomplishes by inserting phase shifters in the appropriate  modes) to find the minimum variance, thus optimizing the entanglement test. Our method is otherwise independent of which values of $\phi_k$  attain that minimum.

\subsection{Fully separable four-mode states}
It is relatively easy to account for all separable and biseparable states in the case of four modes. Let us first calculate $p,q,\Delta(\rho_1)$ for fully separable states. We first consider pure states $|\psi_{s}\rangle$ of the form
\begin{equation}
|\psi_{s}\rangle = \bigotimes\limits^{4}_{i=1}\frac{(|0\rangle + \epsilon_i|1\rangle)}{\sqrt{1+|\epsilon_{i}|^2}},
\end{equation}
for complex parameters $\epsilon_i$.
As argued above, we do not have to consider states with more than a single excitation in any one mode. 
For the pure state $|\psi_{s}\rangle$ a corresponding density matrix can be constructed from $|\psi_{s}\rangle\langle\psi_{s}|$:
\begin{equation}
\rho_{s} = p\rho_0 + q\rho_{1} + r\rho_{\geq2},
\end{equation}
where $$p=\prod\limits^{4}_{i=1}\frac{1}{1+|\epsilon_{i}|^2}$$ and $$q = p\sum\limits^{4}_{i=1}|\epsilon_{i}|^2.$$ 

\begin{figure}
  \includegraphics[scale=0.45]{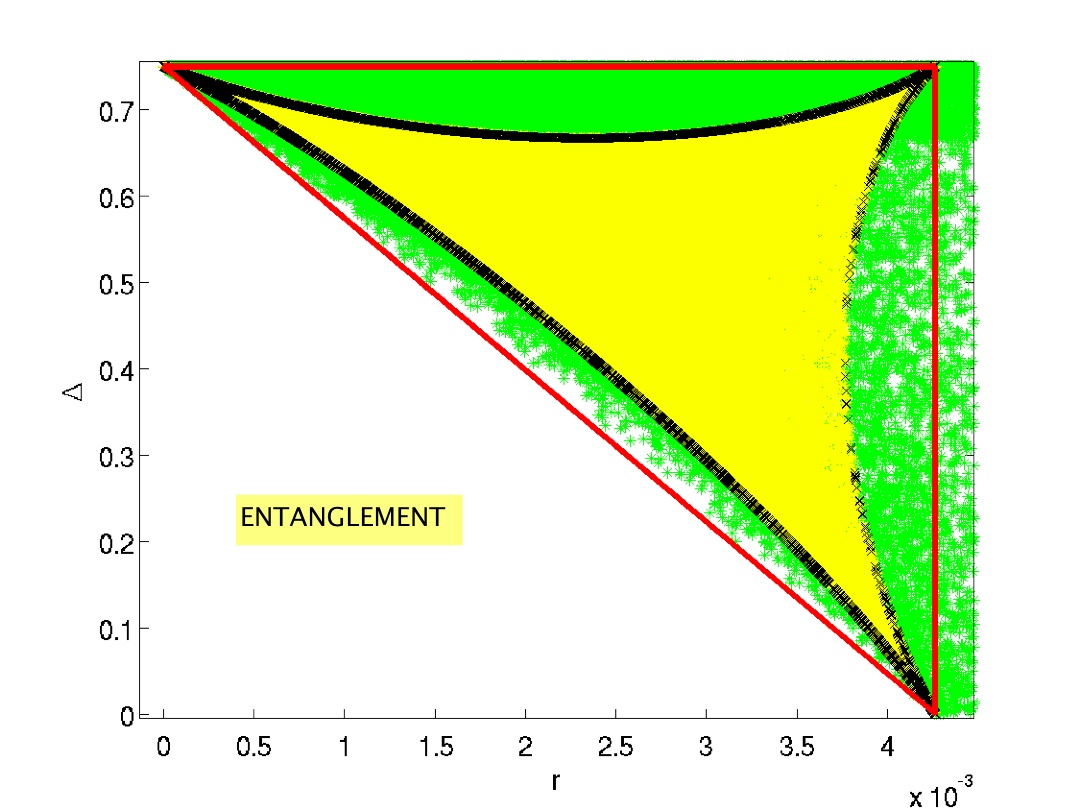}
  \caption{(Color online) Scatter plot (in yellow) of the variance $\Delta(\rho_{s,1})$ of the single-photon part for randomly chosen, pure, fully separable states versus the probability of finding multiple excitations, $r$
  for a fixed single-photon probability $q=0.1$.  The black  crosses are data points for a particular subset of pure states,  attaining the extremum values of the variance for the set of pure states. Also plotted is the variance for randomly chosen {\em mixed} states (in green).  For this particular value of $q$ those values for the variance fall within the convex hull of the graph for pure states (the red line is the convex hull of the black curve). The region below the lowest red line then corresponds to entangled states (as indicated by the word ``ENTANGLEMENT"): but this includes two-mode, three-mode and four-mode entanglement.}\label{WFS}
\end{figure}

We can visualize the set of pure completely separable states, and in particular its border, by plotting values of $\Delta(\rho_{s,1})$ versus  $r$ for a fixed value of the single-excitation probability $q$, by randomly varying over all values of $\epsilon_i$ consistent with that value of $q$. By symmetry, it is clear the minimum variance will be obtained for real parameters. The result is shown in Figure~\ref{WFS}, and we can clearly identify the region of full separability, the lightly shaded area (colored in yellow, online). The minimum value of $\Delta(\rho_{s,1})$ at $r=0$ is $3/4$ which is an agreement with our previous discussion (see Fig.~\ref{Ent}). It is instructive to point out once again that, even though $\Delta(\rho_{s,1})$ approaches zero for sufficiently large values of $r$ [see also the last subsection of this Section], the density matrix $\rho_{s}$ is and remains {\it fully separable} [cf.\ the example mentioned above]. 

Moreover, we managed to find the pure states living on the pure-state boundary, indicated in black in Figure \ref{WFS}. The boundary can be parameterized by two parameters, either $q$ and $r$, or, more simply, by $\epsilon$ and $\tilde{\epsilon}$. Namely, the extremum values of the variance for pure fully separable states turns out to be attained for states of the form
 $$|\psi_{\epsilon,\tilde{\epsilon}}\rangle\propto (|0\rangle+\epsilon |1\rangle)(|0\rangle+\tilde{\epsilon} |1\rangle)^{\otimes 3}.$$
Now one notices the lower border for pure states in Figure \ref{WFS}  is not convex as plotted. This may indicate that points corresponding to certain mixed states may fall below the pure-state boundary. Thus we have also tested randomly chosen mixtures of random pure states as well as mixtures of states on that boundary. And some mixed states (plotted in green) indeed have a smaller variance. Thus, the minimum variance is attained by mixed states in this case, and the correct lower bound is indicated in red. This lower bound coincides with the convex hull of the graph for pure states. 
\subsection{Biseparable states with at most two-mode entanglement}
The next class of states to consider is biseparable states i.e. states that can be described by a density matrix $\rho_{bis} = \sum\limits^{M}_{i=1}\rho^{A}_{i}\otimes\rho^{B}_{i}$. The division into subsystems A and B in the case of four modes has two distinct possibilities -- either system A represents {\em one} of the modes and system B consists of the remaining {\em three} modes or both systems A and B represent {\em two} modes each. We will study the latter case first. 
We represent a pure biseparable state with at most two-mode entanglement by
\begin{eqnarray}
|\psi\rangle_{AB}=
|\psi\rangle_A\otimes |\psi\rangle_B,
\end{eqnarray}
with both two-mode states $|\psi\rangle_k$ for $k=A,B$ of the form
\begin{equation}
|\psi\rangle_k\propto|00\rangle+\epsilon_k|01\rangle+\epsilon'_k|10\rangle,
\end{equation}
where we now included phase factors into the parameters  $\epsilon_k$ and $\epsilon'_k$.
For the same reason as given in the preceding subsection, we do not have to consider contributions from terms with more than a single excitation in any one mode.
The expression for the variance $\Delta$ is symmetric under the interchange of any two modes, and so it is immaterial which two modes constitute system $A$.
We again vary over the complex parameters $\epsilon_k,\epsilon'_k$ for fixed value of the single-photon probability $q=0.1$ to find the set of all pure biseparable states with at most two-mode entanglement, as a function of the multiple-excitation probability $r$. By symmetry the minimum variance is attained for real coefficients. 

\begin{figure}
  \includegraphics[scale=0.45]{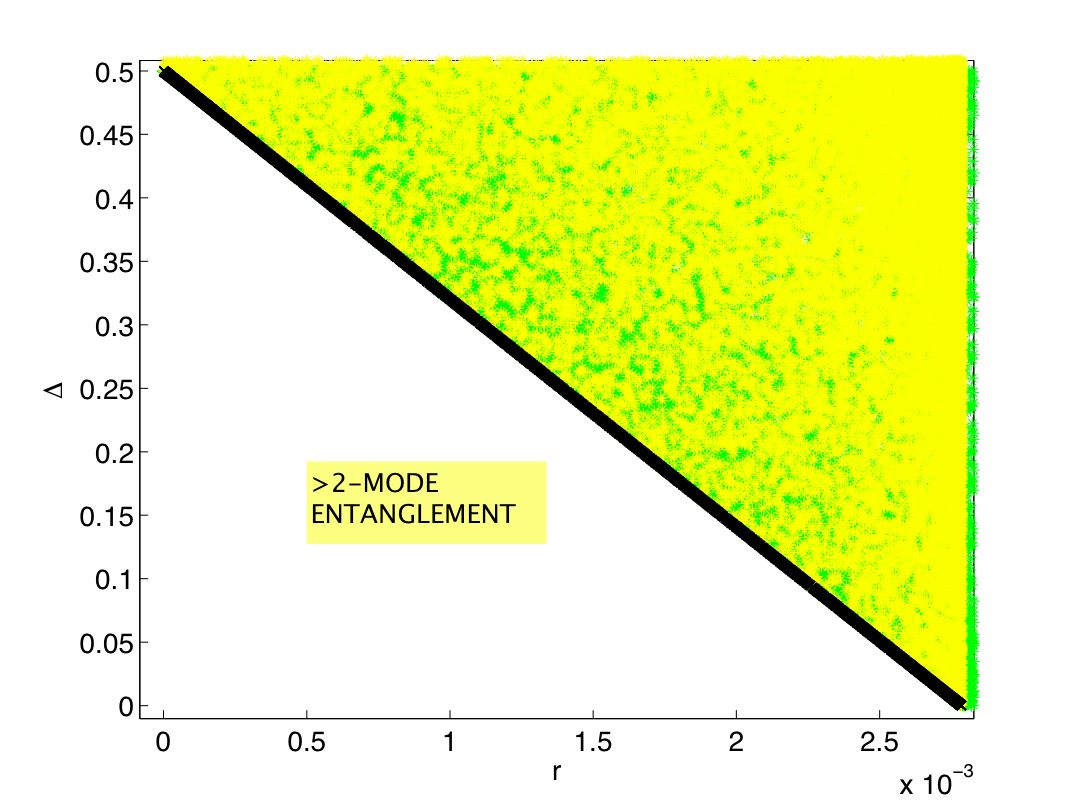}
  \caption{(Color online.) Same as Figure \ref{WFS}, but for pure biseparable states with at most two-mode entanglement. The graph is convex, and points corresponding to mixed states (plotted in green) fall  within the yellow region. The region below the black curve corresponds to at least three-mode entanglement.}\label{W22}
\end{figure}

The lightly-shaded (yellow) area in Figure \ref{W22} then depicts the set containing all biseparable states with at most two-mode entanglement. Indeed, we have checked explicitly that points corresponding to mixed states fall within the shaded region, unlike in the preceding case of fully separable states. The shaded region of Figure \ref{W22} includes that of Figure \ref{WFS}, simply because the set of fully separable states is a subset of the set of states with at most 2-mode entanglement. The minimum value of $\Delta$ at $r=0$ is 1/2, confirming the result from Fig.~\ref{Ent}.
Just as in the preceding subsection we find the pure states living on the boundary. The boundary is again parameterized by two parameters, $\epsilon$ and $\tilde{\epsilon}$. Namely, the minimum variance is attained for biseparable states of the form
 $$|\psi_{\epsilon,\tilde{\epsilon}}\rangle_{AB}\propto (|00\rangle+\epsilon |10\rangle+\epsilon |01\rangle) (|00\rangle+\tilde{\epsilon} |10\rangle+\tilde{\epsilon} |01\rangle).$$
In this case it is straightforward to extract the minimum variance as a function of $q$ and $r$:
\begin{equation}
\Delta_{\min}=1/2-2r(1-q)/q^2+2r^2/q^2,
\end{equation}
which is indeed almost linear in $r$ when $r\ll 1$. Moreover, this boundary is convex. We can rewrite the minimum variance more compactly as
\begin{equation}
\Delta_{\min}=1/2-2rp/q^2.
\end{equation}
\subsection{Biseparable states with three-mode entanglement}
A pure state of the entire four-mode system (with up to two excitations) that has at most three-mode entanglement can be described by the following biseparable vector,
\begin{eqnarray}
|\psi\rangle_{AB}\propto (|0\rangle + \epsilon_{1}|1\rangle)\otimes\nonumber\\
(|000\rangle +\epsilon_{2}|100\rangle + \epsilon_{3}|010\rangle + \epsilon_{4}|001\rangle),
\end{eqnarray}
where we have arbitrarily  chosen the first mode to be the system $A$. The analysis, however, is symmetric with respect to our choice for the system $A$. 
In the second term we do not have to consider states with more than a single photon in system $B$. Although the measurement determining whether there are multiple excitations in the three modes comprising system $B$ is not a local filtering operation in the usual sense, it is local with respect to the bipartite cut $A$ vs $B$, which is the relevant cut in this case.

\begin{figure}
  \includegraphics[scale=0.45]{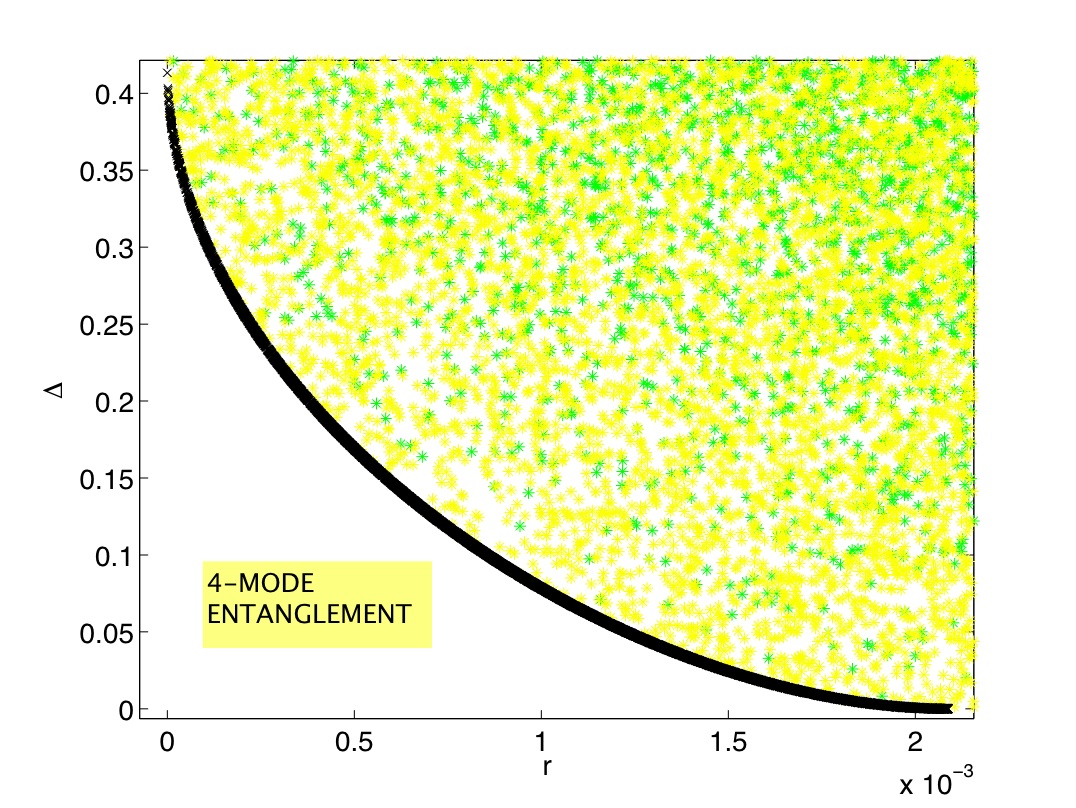}
  \caption{(Color online.) Same as Figure \ref{WFS} ($q=0.1$), but for biseparable states with at most three-mode entanglement. Points corresponding to mixed states fall within the lightly shaded (yellow) region, and are plotted in green. The region below the black curve corresponds to states with genuine four-mode entanglement.}\label{W13}
\end{figure}

The result for $q=0.1$ is plotted in Fig.~\ref{W13}. Like in the case of separable states we observe the existence of points where the values of $\Delta(\rho_1)$ are close to zero. Again we should not misjudge the presence of entanglement in these states,  since the states we are operating with are biseparable by construction. The region for $r$ larger than $\approx 2\times 10^{-3}$ where the minimum variance no longer is a decreasing function of $r$, contains no physical states with smaller variance.

The lightly-shaded (yellow) region depicts the convex set of biseparable states with at most three-mode entanglement, and includes the set of fully separable states, although not necessarily the set of states with {\em two} two-mode entangled states [it does for $q=0.1$]. We have explicitly verified that points corresponding to mixed states (plotted in green) fall within the yellow region. The minimum value of $\Delta$ at $r=0$ is perhaps a little hard to discern, but is indeed equal to 5/12, the value obtained analytically in the preceding Section. The lower boundary
(plotted in black) corresponds to states of the form
\begin{eqnarray}
|\psi\rangle_{AB}\propto (|0\rangle + \tilde{\epsilon}|1\rangle)\otimes\nonumber\\
(|000\rangle +\epsilon|100\rangle + \epsilon|010\rangle + \epsilon|001\rangle),
\end{eqnarray}
with real and positive $\epsilon>\tilde{\epsilon}$.
Since the boundary of minimum variance is the lowest for this type of biseparable states, it is the relevant boundary for the purpose of detecting genuine 4-mode entanglement. For this reason we plot these boundaries for several values of $q$. 
\begin{figure}
  \includegraphics[scale=0.45]{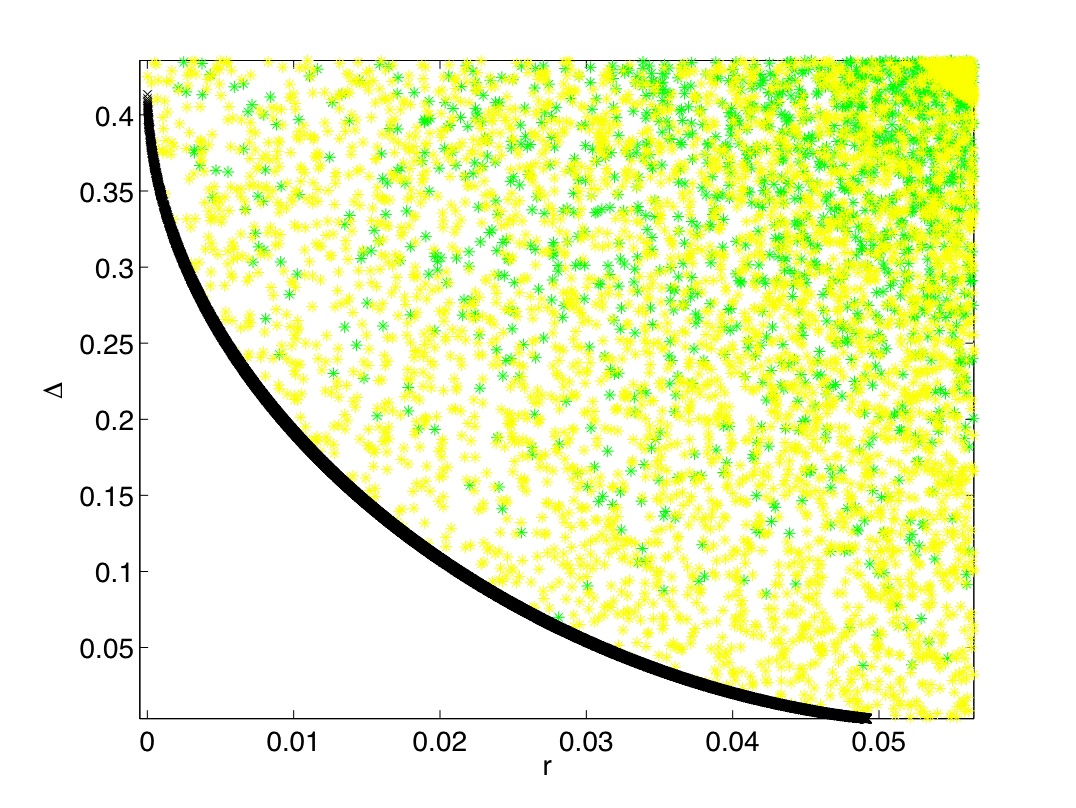}
  \caption{(Color online.)  Same as Fig.~\ref{W13} but for $q=0.4$}\label{q04W13}
\end{figure}
\begin{figure}
  \includegraphics[scale=0.45]{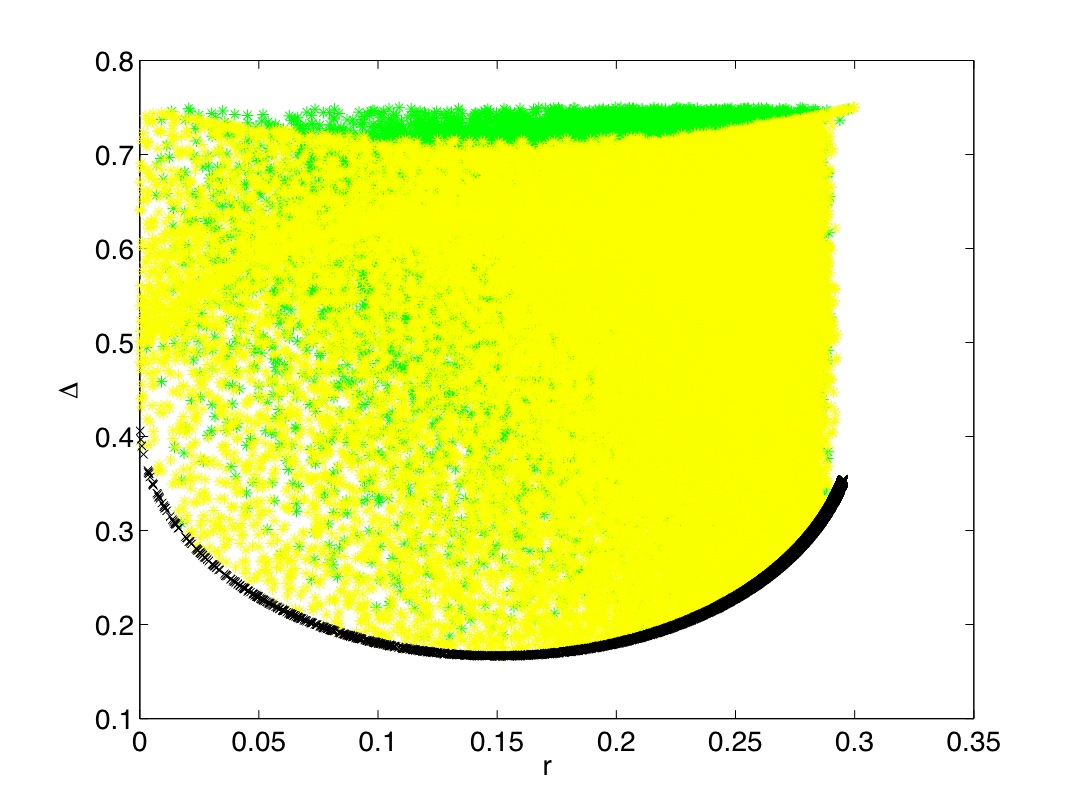}
  \caption{(Color online.) Same as Fig.~\ref{W13} but for $q=0.7$}\label{q07W13}
\end{figure}
\begin{figure}
  \includegraphics[scale=0.45]{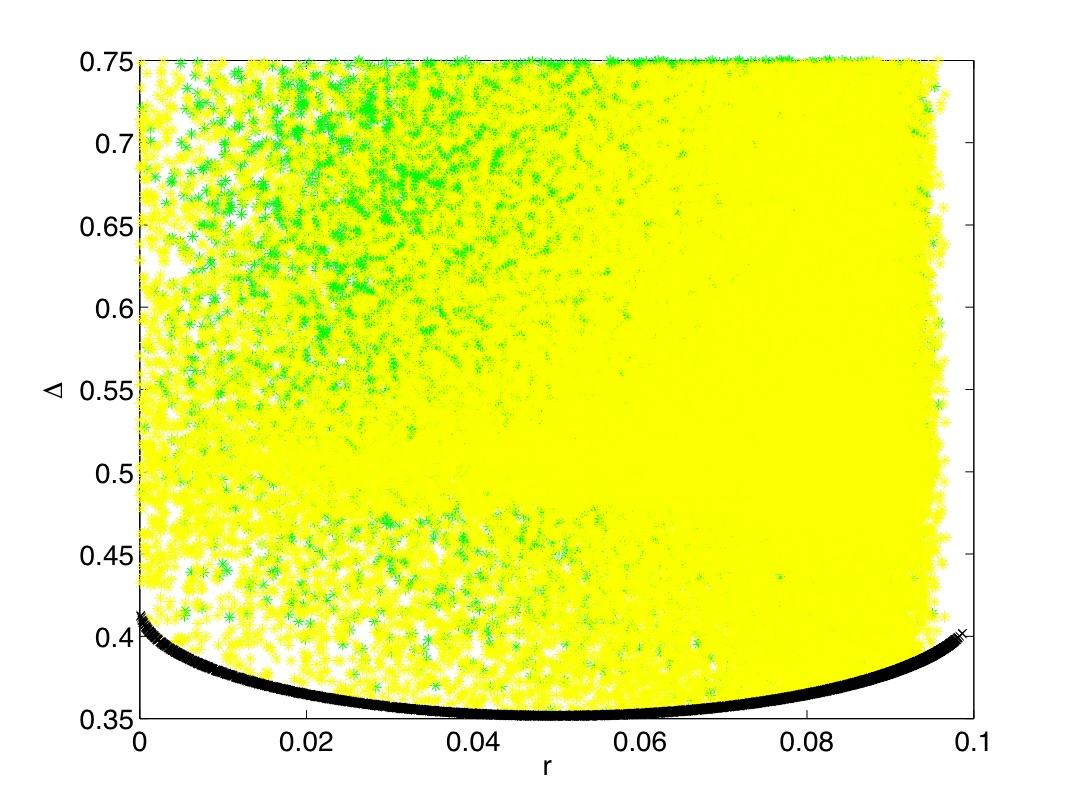}
  \caption{(Color online.)  Same as Fig.~\ref{W13} but for $q=0.9$.}\label{q09W13}
\end{figure}
For increasing values of $q$ the minimum possible variance for 3-mode entangled states increases and reaches the limit of $\max(\Delta_{\min})=5/12$ for $q\rightarrow 1$.
Figures \ref{q04W13}--\ref{q09W13} approach this limit for values $q=0.4$ through $q=0.7$ to $q=0.9$. 
\subsection{Putting it all together}
Based on an exclusion analysis, a practical inseparability criterion can be formulated. In an experiment aimed at detecting a genuinely four-mode entanglement one   measures the values of $r$,  $q$ and $\Delta$. Then one plots, according to the previous considerations, values of  $\Delta$ versus $r$ for all separable and biseparable-state models, feeding in the value of $q$ attained from the experiment. The measured values of  $p$,  $q$ and $\Delta$ are represented by a single point in that plot. If that point lie {\em outside} all three shaded regions of the model plots, the state produced in the experiment must carry genuine four-mode entanglement. Partial conclusions about entanglement can be reached when the point falls outside some and inside other regions.
\begin{figure}
  \includegraphics[scale=0.45]{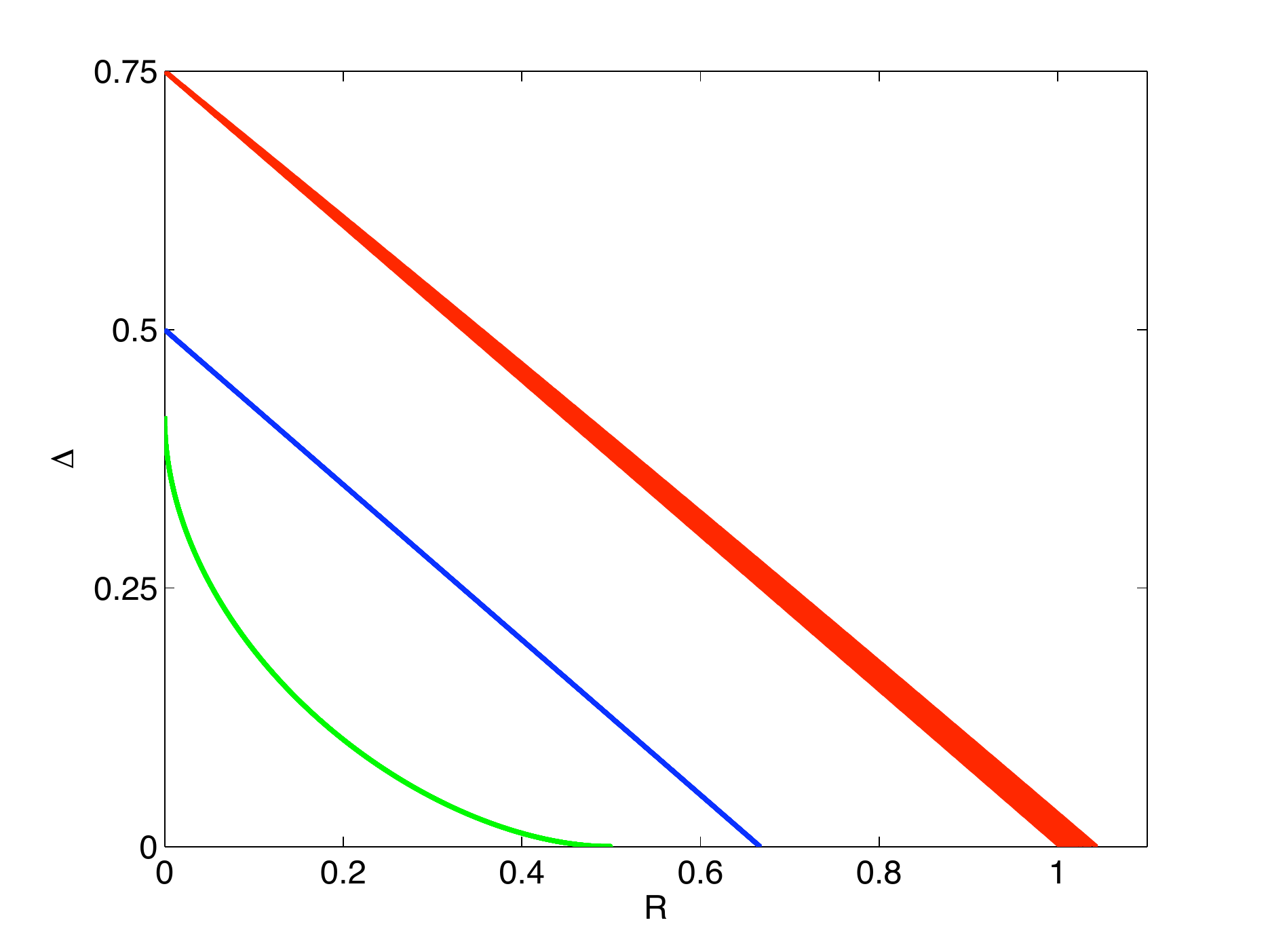}
  \caption{(Color online.) Boundaries for the minimum variance for the three types of biseparable states as functions of $R:=8rp/3q^2$
  for 10 values of $q=0.02,0.04,\ldots 0.2$.  The reason for choosing this particular variable is given in Section \ref{R}.  The lowest-lying (green) curves correspond to 3-party entangled states, the highest-lying (red) curves correspond to fully separable states, the middle (blue) curves  correspond to biseparable states with 2 -mode entanglement. The variance depends only weakly on $q$ for the red curves, and is independent of $q$ for the green and blue curves. }\label{boundaries}
\end{figure}
In particular, if the measurement point lies outside the shaded region of Fig.~\ref{WFS}, but inside the shaded regions of Figures \ref{W22} and \ref{W13}, one can only conclude one has an entangled state, but it could be merely two-mode entangled. If the point falls outside the shaded regions of both Figures \ref{WFS} and \ref{W22}, but inside the shaded region of Figure \ref{W13}, one has at least three-mode entanglement. Of course, if the point falls inside the shaded region of Figure \ref{WFS}, no firm conclusion can be reached about entanglement, as there is a fully separable state consistent with one's values for $p,q,\Delta$.

We plot the three  minimum-variance boundaries for different small values of $q$ as a function of a scaled variable $R:=8rp/3q^2$ (see the next subsection for an explanation for this choice of variable) in Fig.~\ref{boundaries}. One sees the boundaries depend only weakly on that parameter. 
\subsection{Some necessary conditions for entanglement}\label{R}
 Let us finally consider the conditions on entanglement
in the simple situation where the variance $\Delta(\rho_1)$ vanishes and where $q$ is not too large. We consider the same three classes of unentangled states as before. 
\begin{enumerate}
\item Fully separable states with $\Delta(\rho_1)=0$ must be of the form $(|0\rangle+\epsilon|1\rangle)^{\otimes 4}$. For such states, the point at which the variance is zero is characterized by
\begin{equation}\label{R2}
\frac{8}{3}\frac{rp}{q^2}=1+\frac{q}{6p}+\frac{q^2}{96 p}
\end{equation}
For small values of $q$, we can give the approximate relation, valid for fully separable states:
\[
R\geq 1,
\]
with $R=8rp/3q^2$ being the quantity appearing on the lhs of (\ref{R2}).
A necessary (although not sufficient) condition for any type of entanglement is then simply
\[
R<1.
\]
For Figure \ref{WFS}, in which we took $q=0.1$, this places a strict upper limit on $r$ of $r<4.125\times 10^{-3}$ for entanglement to be detectable through $\Delta$.
\item Biseparable states with $\Delta(\rho_1)=0$ and at most two-mode entanglement must be of the form $(|00\rangle+\epsilon|01\rangle+\epsilon|10\rangle)^{\otimes 2}$. For such states, the boundary of zero variance is at $R=2/3$, and hence all biseparable states satisfy
\[
R\geq  \frac{2}{3}.
\]
For Figure \ref{W22}, in which $q=0.1$, this places a strict upper limit on $r$ of $r<2.75\times 10^{-3}$ for entanglement involving at least three modes to be detectable through $\Delta$.
\item Biseparable states with $\Delta(\rho_1)=0$ and at most three-mode entanglement must be of the form $(|0\rangle+\epsilon|1\rangle)\otimes(|000\rangle+\epsilon|001\rangle+\epsilon|010\rangle+\epsilon|100\rangle$. For such states, we similarly derive 
\[
R\geq \frac{1}{2}.
\]
For Figure \ref{W13}, in which $q=0.1$, this places an upper limit on $r$ of $r<2.06\times 10^{-3}$ for entanglement to be detectable through $\Delta$.
\end{enumerate}
In order to demonstrate genuine four-mode entanglement one must violate all of these conditions. That is, one must violate the strongest of these conditions, and hence one must have
\begin{eqnarray}\label{p2q}
R<\frac{1}{2}.
\end{eqnarray}
This condition for four-mode entanglement is necessary but not sufficient for nonzero values of $\Delta(\rho_1)$.
The form of the conditions also indicates why the scaled variable $R$, used in Fig.~\ref{boundaries}, is a useful quantity for small $q$ for fully separable states, and for biseparable states irrespective of the value of $q$.
\section{Losses and asymmetries }\label{losses}
So far we have assumed  that the variance measurement device is ideal: beamsplitters (see Fig. \ref{setup}) were assumed lossless and perfectly balanced, and detectors were perfect. In this subsection we relax those conditions and describe the modifications necessary to include these imperfections. First, we consider the effect of imbalanced beamsplitters.
\subsection{Imbalanced beamsplitters}
Suppose, then, we have the same setup as depicted in Fig. \ref{setup}, but with the 4 beamsplitters having reflection and transmission probabilities $|t_k|^2$ and $|r_k|^2$ not necessarily equal to 1/2. Consider one output mode, say the top one. 
There is one path a photon can take from input mode 1 to reach the top output mode: it has to reflect off of beamsplitter 1 and it has to reflect off of beamsplitter 3. The amplitude for that path is then $r_1r_3$ in terms of the reflection amplitudes of beamsplitters 1 and 3. Here we ignore phase factors due to propagation (they can be trivially inserted in the end).
Similarly, a photon from input mode 2 can reach the top output mode along just one path, with amplitude $t_1r_3$.
Writing down the amplitudes for photons starting in input modes 3 and 4 shows that
a photo-detection at the top output mode projects onto the (input) state
\begin{equation}
|\tilde{W}_1\rangle=r_1r_3|1000\rangle+t_1r_3|0100\rangle+
r_2t_3|0010\rangle+t_2t_3|0001\rangle.
\end{equation}
This is a properly normalized state, even if the beamsplitters are not balanced. The normalization follows from the relation $|r_k|^2+|t_k|^2=1$ for lossless beamsplitters.

We can similarly write down the states onto which one projects if detecting a photon in one of the remaining output modes:
\begin{eqnarray}
|\tilde{W}_2\rangle&=&t_1r_4|1000\rangle+r_1r_4|0100\rangle+
t_2t_4|0010\rangle\nonumber\\ &&+r_2t_4|0001\rangle,\\
|\tilde{W}_3\rangle&=&r_1t_3|1000\rangle+t_1t_3|0100\rangle+
r_2r_3|0010\rangle+\nonumber\\ && t_2r_3|0001\rangle,\\
|\tilde{W}_4\rangle&=&t_1t_4|1000\rangle+r_1t_4|0100\rangle+
t_2r_4|0010\rangle\nonumber\\ && +r_2r_4|0001\rangle.
\end{eqnarray}
These states, too, are normalized. Moreover, the 4 states are all orthogonal, as follows from the (unitarity) relation $t_k^*r_k+t_kr_k^*=0.$
One can still calculate the variance of photodetector counts, using the modified projectors onto the $\tilde{W}$ states, but that variance will not give as much information as in the balanced case about four-mode entanglement.
For example, consider the extreme case of a mirror replacing beamsplitter 4: that is, assume now that $r_4=1$ and $t_4=0$. Then states $|\tilde{W}_2\rangle$ and
$|\tilde{W}_4\rangle$ are no longer four-mode entangled states, but only two-mode entangled states. Thus, certain two-mode entangle states would give rise to a zero variance in this extreme case.

This implies that even if one's experiment cannot use perfect 50/50 beamsplitters, one should at least try to make them as balanced as possible.
In such cases one needs in general a lower variance $\Delta$ than in the ideal balanced case
to conclude one has four-mode entanglement.
\subsection{Losses}
Now consider losses. We can model linear losses (both propagation losses, and inefficiencies of the photodetectors) by imagining lossless paths but with additional beamsplitters reflecting away some portion of the light in the lossy paths.
The output of those additional beamsplitters does not lead to the output detectors, but to other (unmonitored) output modes. The overall transformation from input to output is still unitary, which implies there must also be additional input modes (just as many as there are unmonitored output modes). A photodetection in one of the desired output modes projects onto a set of orthonormal states on the larger Hilbert space of all input modes. If we write down the projections of those states onto the 4 input modes of interest, we will end up with subnormalized states. For example, considering for the moment (see the next subsection where we take into account multiple excitations) only states with exactly one photon, a detection in the top output mode projects onto the state
\begin{eqnarray}
|\tilde{W}'_1\rangle=T_{11}r_1r_3|1000\rangle
+T_{21}t_1r_3|0100\rangle\nonumber\\
+
T_{31}r_2t_3|0010\rangle+T_{41}t_2t_3|0001\rangle
\end{eqnarray}
where the transmission amplitude $T_{k1}$ is the product of all loss amplitudes encountered by a photon propagating from input $k$ to the top output detector (this includes the inefficiency of the detector).

The variance we are interested in is conditioned on detecting (at least) one photon in the desired output modes. Once we detect a photon in the top mode, we do renormalize the state $|\tilde{W}'_1\rangle$ and project onto:
\begin{eqnarray}
|\tilde{W}''_1\rangle=\frac{|\tilde{W}'_1\rangle}{\sqrt{\langle \tilde{W}'_1|\tilde{W}'_1\rangle}}.
\end{eqnarray} 
The 4 states onto which we project conditionally, $|\tilde{W}''_k\rangle$ for $k=1\ldots 4$ are, therefore, properly normalized, but they are not orthogonal, unless all losses are balanced. i.e., if $T_{lk}=$ constant for all $l,k=1\ldots 4$. 

Again, we still can use a variance based on the modified nonorthogonal projectors, but that variance will give less information than in the ideal lossless balanced case. For instance, if all photodetectors but one are completely inefficient and never detect any photon, the variance would be zero for any input state. Thus, in an actual experiment one would have to make the losses as balanced as possible in order for the variance to contain as much information about four-mode entanglement as possible. 
Of course, one would also like to limit the size of the  losses for various different reasons.

\begin{figure}
  \includegraphics[scale=0.45]{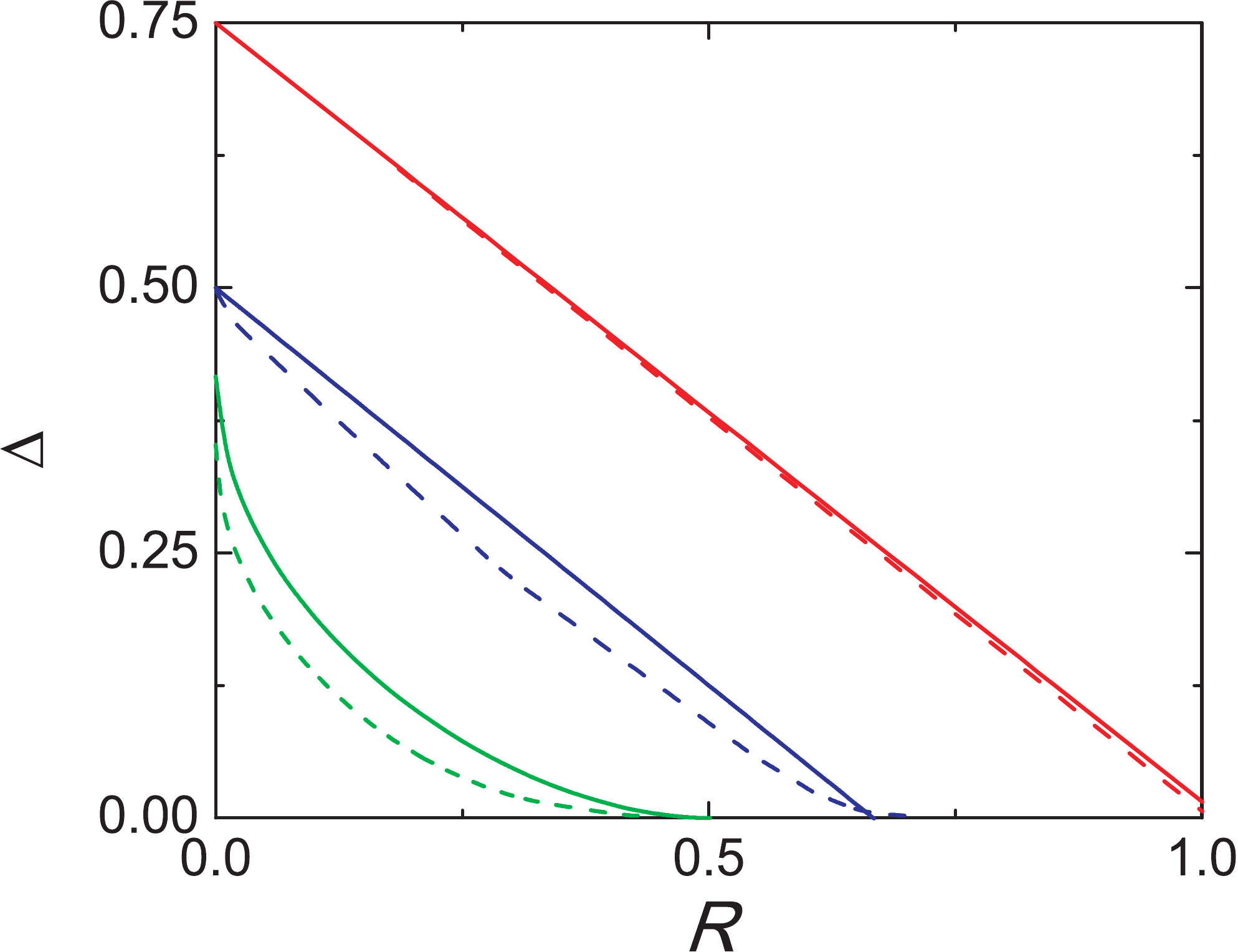}
  \caption{(Color online.) Minimum variance curves for
  the case where all beamsplitters are 55/45 (dashed lines) rather than 50/50 (solid lines).}\label{balance}
\end{figure}

\begin{figure}
  \includegraphics[scale=0.45]{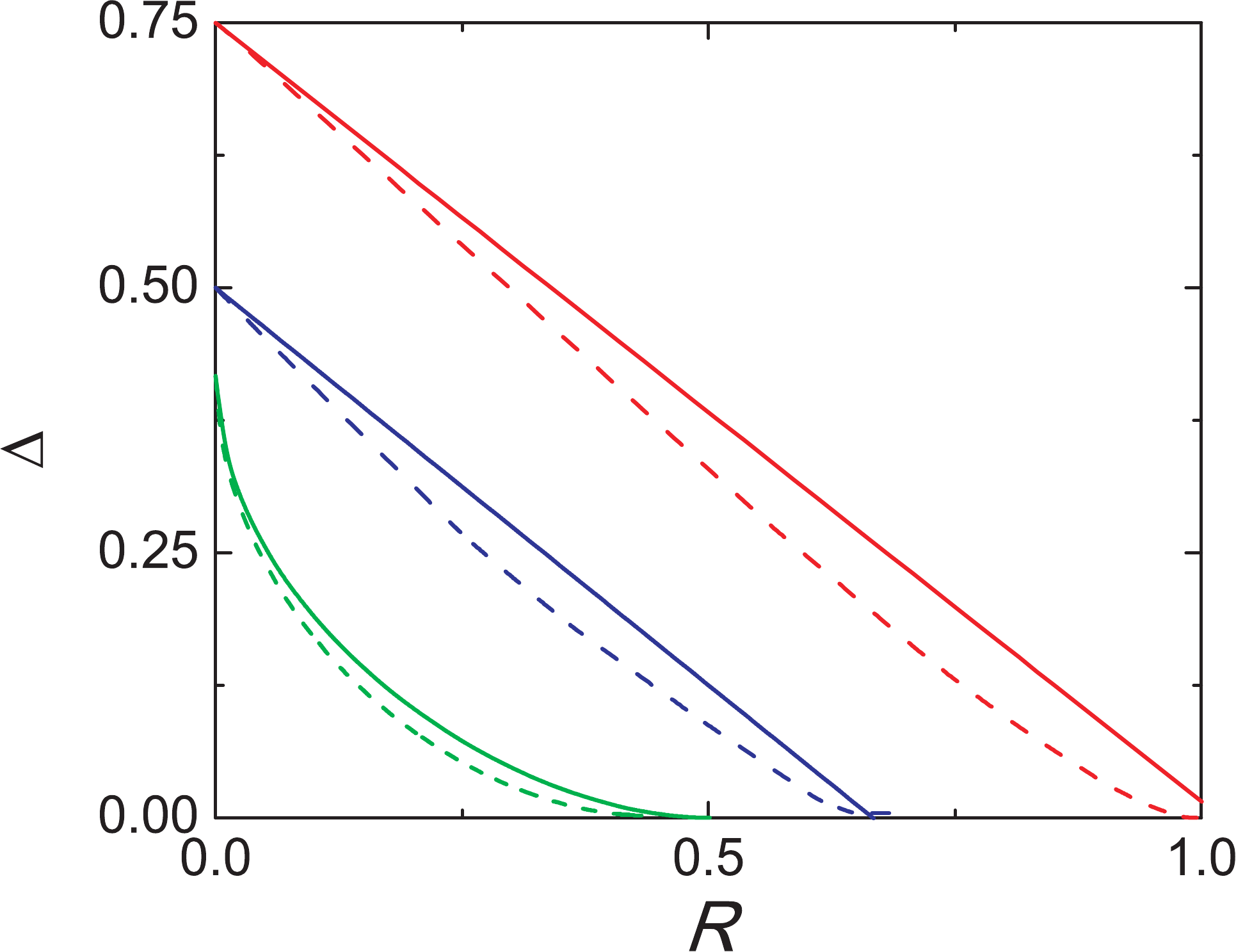}
  \caption{(Color online.) Minimum variance curves for
  the case where there is one lossy path with transmission probability of 60\% (a typical parameter) (dashed lines),  compared to the ideal lossless case (solid lines).}\label{loss}
\end{figure}

\begin{figure}
  \includegraphics[scale=0.45]{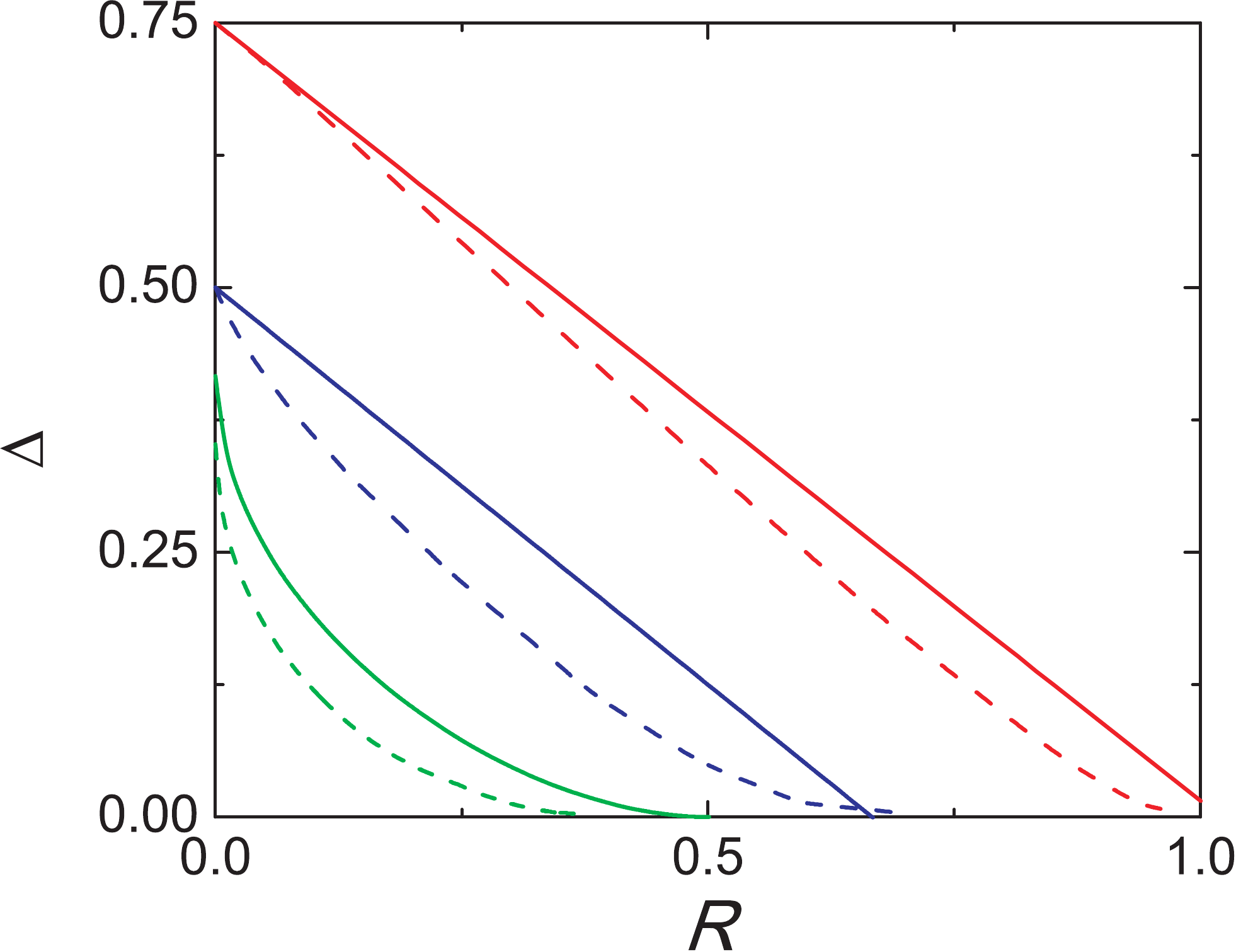}
  \caption{(Color online.) Minimum variance curves for
  the case where there is one lossy path with a transmission probability of 60\% and unbalanced 55/45 beamsplitters (dashed lines), compared to the ideal lossless and balanced case }\label{lossbalance}
\end{figure}

With the new projectors onto the nonorthogonal states $|\tilde{W}''_k\rangle$ in hand, we can perform the same calculations as we did in the ideal case: find the minimum variance consistent with unentangled input states, input states with two-mode entanglement, and input states with three-mode entanglement, respectively,  for fixed values of $q$ and $r$. 
We display three illustrative examples (for $q=0.1$): in Fig.~\ref{balance}
we assume no losses but unbalanced beamsplitters,
in Fig.~\ref{loss} we assume losses, but balanced beamsplitters, and in Fig.~\ref{lossbalance} we show the effects of the combination of both losses and imbalances.
All figures show the tendency of the minimum variance to decrease compared to the ideal lossless and balanced case.

\subsection{Measured variance vs $\Delta(\rho_1)$}
In the presence of losses the measured variance, $\Delta_m$, is not just due to the single-excitation part,  but  from the multi-excitation part of the input state as well. Thus, the measured variance has to be corrected (upwards in fact) in order to find an estimate for the variance $\Delta(\rho_1)$ due to the single-excitation part, because that is the quantity we used above to detect entanglement. 

We discuss a simple case (balanced losses throughout the system and the use of non-number resolving threshold detectors \footnote{The effect of imbalanced beamsplitters in the presence of balanced losses is easily included in this calculation. The final bound,  including imbalanced losses and imbalanced beamsplitters, has the same form   (\ref{boundV}) with the same expression for $q_1=p_1'q_1'/Q$.})
where we find we simply have to multiply the measured variance with a factor $c>1$ to obtain an upper bound on $\Delta$. That is, the variance $\Delta$ is upper-bounded by $c\Delta_m$. We now evaluate $c$.

Consider the propagation of the purported experimental state (here we change notation to make it easier to keep track of the meaning of all symbols)
\begin{equation}\label{state}
\rho_W=p_0 \rho_{0}+p_1 \rho_{1}+p_{\geq 2} \rho_{\geq2}.
\end{equation}
Under balanced losses (which can be characterized by a single transmission efficiency $|T|^2$)
this state
transforms to $\rho_{T}$ where
\begin{equation}\label{lossstate}
\rho_{T}=p^{\prime}_{0}\rho_{0}+p^{\prime}_{1}(q^{\prime}_{1} \rho^{(1)}_{1}+(1-q^{\prime}_{1}) \rho^{(2)}_{1})+p^{\prime}_{\geq 2}\rho^{(2)}_{\geq2}.
\end{equation}
Here, $\rho^{(i)}_{1}$ is the 1-photon subspace of $\rho^{(r)}_{|T|^2}$
originating from the $i$-photon $\rho_{i}$ subspace of $\rho^{(r)}_{W}$
for $i\in\{1,2\}$, and $\rho^{(2)}_{\geq 2}$ is the 2-photon subspace
after the transmission. To the leading order of $\rho^{(2)}_{\geq 2}$
(neglecting 3-photon and 4-photon subspaces), $\{p^{\prime}_{0},p^{\prime}_{1},p^{\prime}_{2}\}$
are
\begin{equation}
p^{\prime}_{0}=p_0+(1-|T|^2)p_1+(1-|T|^2)^{2} p_2
\end{equation}
\begin{equation}
p^{\prime}_{1}=|T|^2 p_{1} +2 |T|^2 (1-|T|^2) p_2
\end{equation}
\begin{equation}
p^{\prime}_{2}=|T|^4 p_2,
\end{equation}
and  $q^{\prime}_{1}$ is given as
\begin{equation}
q^{\prime}_{1}=\frac{|T|^2 p_{1}}{|T|^2 p_{1} +2 |T|^2 (1-|T|^2) p_2}.
\end{equation}

Thus, if we denote by $P_k$  the normalized probability of the detector  (assumed to be non-number resolving) in
mode $k$ finding (at least) one photon , then we have
\begin{equation}
P_k=\frac{p^{\prime}_{1} q^{\prime}_{1}}{Q}P^{(1)}_{1,k}
+\frac{p^{\prime}_{1} (1-q^{\prime}_{1})}{Q}P^{(1)}_{2,k}
+\frac{p^{\prime}_2}{Q}P^{(2)}_{2,k},
\end{equation}
where $Q=p^{\prime}_{1}+p^{\prime}_{2}$.
Here, $P^{(1)}_{1,k}$ ($P^{(1)}_{2,k}$) is
the probability of a 1-photon in output mode $k$
originating from the 1(2)-photon subspace $\rho_{1}$ ($\rho_{2}$),
and $P^{(2)}_{2,k}$ is the probability of 2-photon in output mode $k$.
To be conservative (for our purposes of finding a {\em sufficient} condition for entanglement), we assume that
the two photons are directed towards {\em one} detector at a time
so that we cannot distinguish $P^{(1)}_{1,k}$ from $P^{(2)}_{2,k}$.
By denoting $q_{1}=p^{\prime}_{1} q^{\prime}_{1}/Q$
as the probability of detecting desired events and $X_{k}$ as the normalized probability of detecting undesired events (that is, $P^{(1)}_{2,k}$ and $P^{(2)}_{2,k}$), we get
\begin{equation}
P_k=q_1 P^{(1)}_{1,k} + (1-q_1) X_{k}.
\end{equation}
The measured variance $\Delta_m$ is given as
\begin{equation}
\Delta_m=1-\displaystyle\sum_k P_k^2.
\end{equation}
On the other hand, the 1-photon variance $\Delta$ 
is defined as 
\begin{equation}
\Delta(\rho_1)=1-\displaystyle\sum_k (P^{(1)}_{1,k})^2.
\end{equation}
As a conservative correction to $\Delta_{m}$,
we assume that the unwanted events ($X_k$)
are all directed towards the output mode $j$ which contains the maximum 1-photon probability $P^{(1)}_{1,j}$
(i.e. $X_j=1$ and  $X_k=0$ for $k\neq j$). This way, the measured variance is {\em lower} than the variance $\Delta$. Thus our conservative bound gives then
\begin{equation}
\Delta_m=1-(q_1 P^{(1)}_{1,j} + (1-q_1))^2-q_1^2 \displaystyle\sum_{k\neq j} (P^{(1)}_{1,k})^2.
\end{equation}
Using the inequality $2(1-P^{(1)}_{1,j})\geq (1+P^{(1)}_{1,j})(1-P^{(1)}_{1,j})=1-(P^{(1)}_{1,j})^2$,
we obtain
\begin{equation}\label{boundV}
\Delta_m\geq q_1 [\Delta+(1-q_1)\displaystyle\sum_{k\neq j} (P^{(1)}_{1,k})^2] \geq q_1 \Delta.
\end{equation}
Therefore, we obtain a correction factor of
\begin{equation}
c=\frac{1}{q_{1}}
\end{equation}
where
\begin{equation}\label{balanceeq}
q_{1}=\frac{p^{\prime}_{1} q^{\prime}_{1}}{Q}=\frac{p_{1}}{p_{1}+(2-|T|^2)p_{\geq 2}}.
\end{equation}
In the limit of $p_{0}\approx 1$,
the correction factor becomes
\begin{equation}
c\approx 1+\frac{3}{8} (2-|T|^2)p_{1} R.
\end{equation}

 \section{Summary and Discussion}
 We demonstrated how to verify $N$-party entanglement
 of W states or states lying close to W states, in the case quantum information is encoded in the number of excitations per mode.
 Our method takes into account the presence of the vacuum state, as well as multiple excitations; moreover, it takes into account losses during the verification measurements, as well as imperfect beamsplitters. The method applies to any number of modes, but we focused on four modes for illustrative purposes, as the method was applied in an actual experiment \cite{exp} to four modes. A relatively straightforward set of measurements allows one, in that case, to distinguish genuine four-party entanglement from three-party entanglement, which in turn can be distinguished from two-party entanglement and fully separable states. One must obtain estimates of three parameters: a variance $\Delta$ determined by the single-photon part of the state, the single-photon probability $q$, as well as the multi-photon probability $r$. For example, the simple condition of Eq. (\ref{p2q}) is a necessary condition for genuine four-party entanglement (where our definition of genuine multi-partite entanglement is more severe than usual) which involves only $r$ and $q$. To obtain sufficient conditions one must also include the value of $\Delta$ in the analysis.

The measurement of $\Delta$ combines the various modes by simple beamsplitters, and is thus a nonlocal measurement. In this way one does {\em not} need local oscillators, which one would need if the entanglement verification method used local measurements only  \cite{stroud}. In our case  the modes interfere with each other, rather than  with external reference beams. Thus, our method cannot be applied to eliminate local hidden variable models (through Bell inequalities, for example, in the bi-partite case), but it can be applied to verifying entanglement, which is a very different beast indeed \cite{vanEnkKimbleLutkenhaus}.
 \section*{Acknowledgments}
PL and SJvE thank Michael Raymer for useful discussions. This work was supported by NSF, IARPA, and ARO.

\end{document}